\pdfoutput=1 
\documentclass{ptapap}
\usepackage{color}
\usepackage{soul}
\usepackage{amsmath}

\usepackage{amssymb}

\author{Ayush Moharana}[CAMKT]
\author{Krzysztof G. He{\l}miniak}[CAMKT]
\author{Fr\'ed\'eric Marcadon}[CAMKT]
\author{Tilaksingh Pawar}[CAMKT]
\author{Maciej Konacki}[CAMK]
\affil[CAMKT]{Nicolaus Copernicus Astronomical Center, Polish Academy of Sciences, Rabia\'nska 8, 87--100 Toru\'n, Poland}
\affil[CAMK]{Nicolaus Copernicus Astronomical Center, Polish Academy of Sciences, Bartycka 18, 00--716 Warsaw, Poland}

\title{Evolution and Dynamics of Tight Triple Systems}

\begin{document}

\maketitle

\begin{abstract}

Tight Triple Systems have stars in a hierarchical configuration with a
third star orbiting the inner binary with a period of fewer than 1000
days. Such systems are important for understanding the formation and
evolution of stars in multiple systems. Having a detached eclipsing
binary (DEB) as one of its components allows us to obtain precise
stellar and orbital parameters of these systems. We discuss the process to obtain accurate
parameters of these systems using high-resolution spectroscopy, radial velocity
measurements, and precise space-based photometry. This
 enables us to have a 3D geometrical picture as well as the
metallicity, age, and evolutionary status of these systems.
\end{abstract}

\section{Introduction}

Hierarchical triple systems have the third star at a greater distance compared
to the inner binary separation. Most of these systems have long outer periods and therefore their
dynamics can have timescales of decades or centuries. Meanwhile, a subset
of these triples, called Tight Triple Systems (TTS) have an outer orbit period
of fewer than 1000 days (\citealt{hajdu2017}) and hence their dynamics can be observed in
timescales of years. Detached eclipsing binaries (DEBs) are the source of the
most accurate (<1\%) stellar parameters which are robust and independent of different models and methods (\citealt{maxted2020,korthmoh}). Therefore, TTS with DEB
would serve as an ideal system to extract accurate stellar parameters. Our study probes a sample of 20-25 TTS (Fig.1) which are spectroscopically
double-lined (SB2) or triple-lined (SB3) systems. The systems were extracted from literature, the
spectra database of the \textit{CRÈME} project (P.I.: K.G. Hełminiak) and the photometric
eclipse timing campaign of the \textit{Solaris} project (P.I.: M. Konacki).

\section{Methodology}
\begin{figure}
    \centering
    \includegraphics[width=0.48\textwidth]{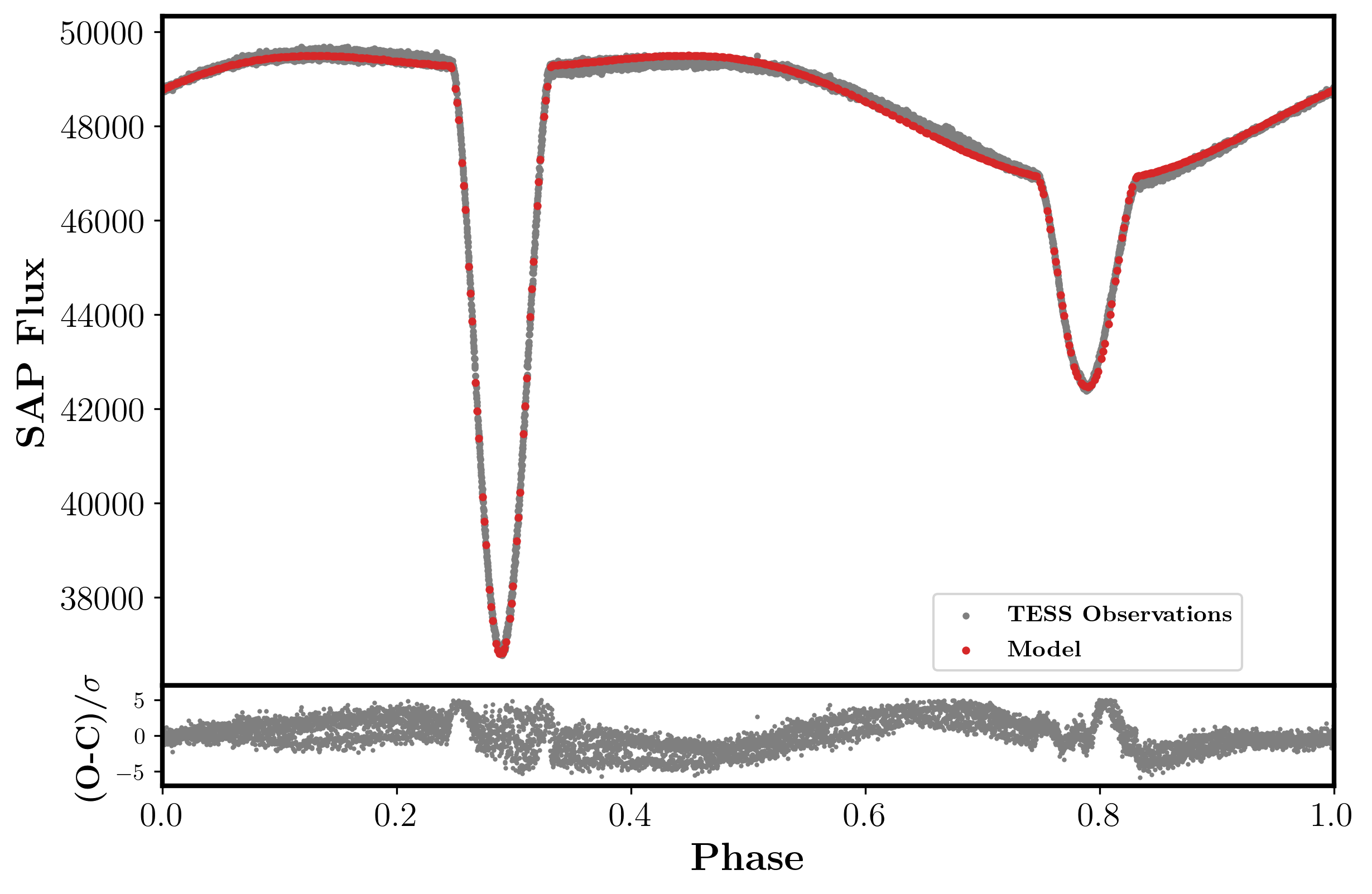}
    \includegraphics[width=0.39\textwidth]{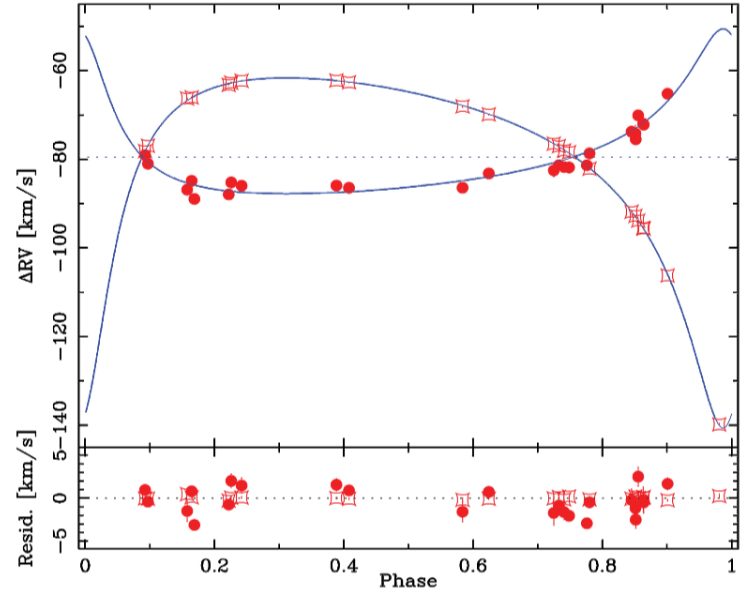}
    \caption{(Left) Phase-folded TESS light curve (grey) with best-fit \textsc{PHOEBE2.3} model (red) for BD+44 5528 (SB3 system). (Right) Phase-folded radial velocities of centre-of-mass of the binary (filled circles) and the tertiary (hollow boxes). Best fit \textsc{V2FIT} model is in blue.}
    \label{fig:method}
\end{figure}

We obtained TESS and/or Kepler observations for our targets from the the
Mikulski Archive for Space Telescopes (MAST). \textsc{PHOEBE2.3} (\citealt{conroy2020}) eclipsing binary modelling code was used for the light curve modelling. The new version of code offers advantages of modelling stellar
spots (Fig.\ref{fig:method}: Left). The errors of the parameters were calculated using
MCMC sampling through the code. The radial velocities were extracted from spectra taken from the \textit{CRÈME} project. We used the \textsc{TODCOR} algorithm to extract radial velocities which were modelled with the \textsc{V2FIT} code (\citealt{v2fit}). The centre-of-mass velocity of the inner-binary (Fig. \ref{fig:method}: Right) was then modelled along with with the tertiary radial velocity (if any). The total mass of the inner binary was used to get the
mass estimate of the tertiary.
\section{Evolution and Dynamics}
\begin{figure}
    \centering
    \includegraphics[width=0.47\textwidth]{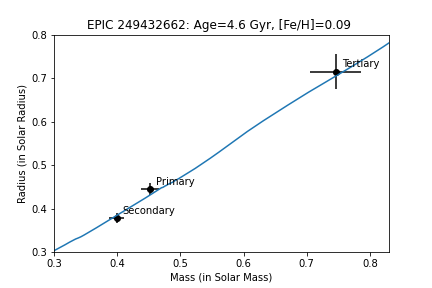}
    \includegraphics[width=0.48\textwidth]{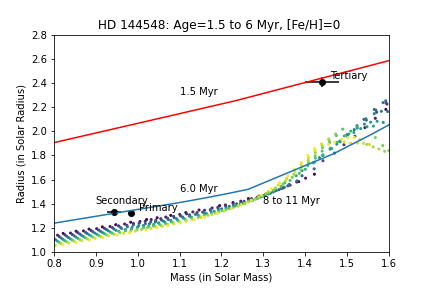}
    \caption{(Left) MESA isochrone in comparison to the parameters obtained for system EPIC 249432662 obtained by \cite{2019borkophotdyn}. All the three stars are of the same age and metallicity, hence form a ``clean" system. (Right) Parameters of HD 144548 obtained by \cite{uscorEB} against different MESA isochrones. The binary stars and tertiary disagree on age given they have the same metallicity. The shaded space corresponds to the age estimate of the OB association in Upper Scorpius to which the system belongs.}
    \label{fig:evotests}
\end{figure}

The measurements of mass and radius give us an opportunity to test evolution scenarios for triple stars. Usually for TTS, we see no interactions between the stars themselves (unless their orbits are very dynamic). This would mean that all three stars in a TTS will evolve separately. Therefore we
expect similar metallicity and age for the three stars in this “clean” system (Fig.\ref{fig:evotests}: Left). Assuming a “clean” system, we can estimate a range for the
radius of the tertiary from the mass-radius relation obtained from the isochrone fitting using the binary. But sometimes this “clean” system assumption fails and all the three stars do not follow the same
isochrone (Fig.\ref{fig:evotests}: Right). This calls for additional probes by independent estimates of metallicity and other relevant stellar parameters (e.g. from spectral analysis of disentangled spectra).

\begin{figure}
    \centering
    \includegraphics[width=0.7\textwidth]{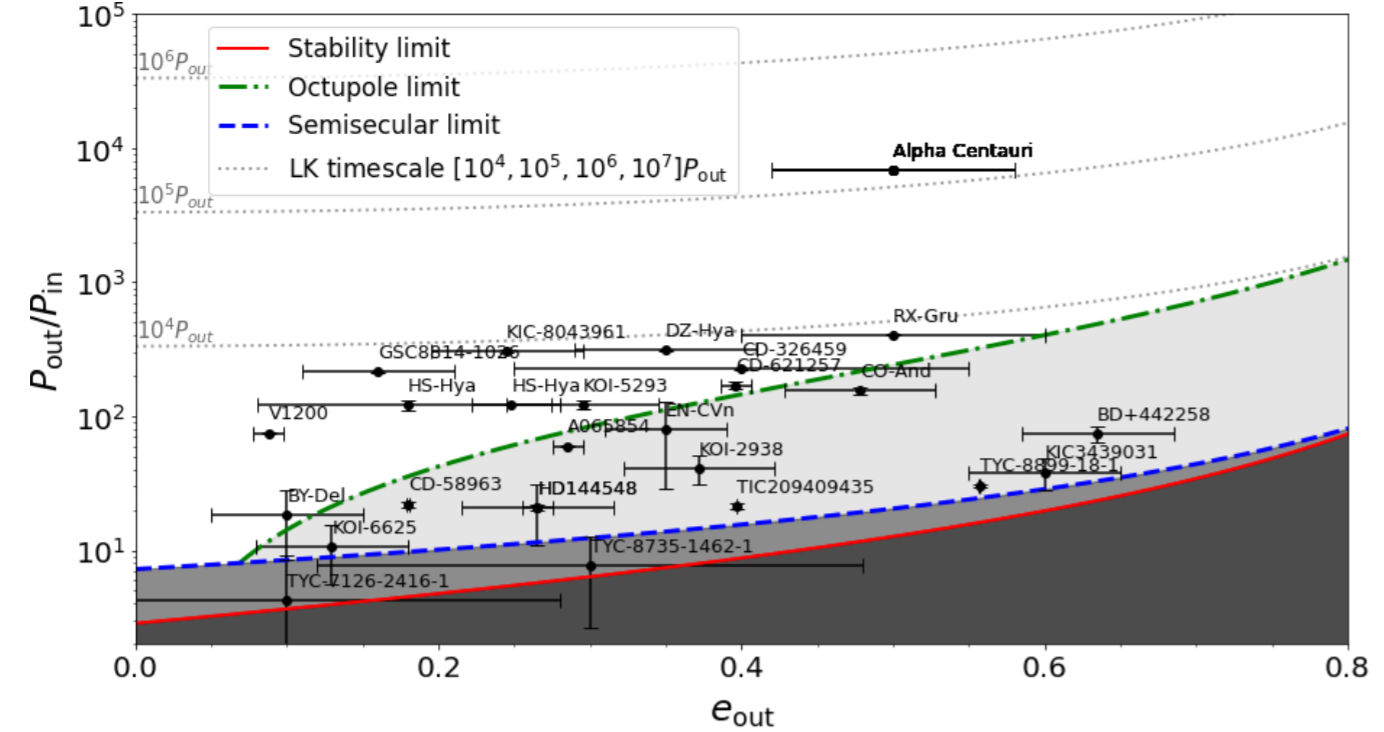}
    \caption{The plot is a visualisation of parameter space of different dynamical regimes in a triple system, based on \cite{toonen2020}, where $e_{out}$ is outer orbit eccentricity and period ratio of outer and inner orbits is $P_{out}/P_{in}$ The proposed systems are over-plotted to this configuration. Alpha Centauri triple system is plotted as a reference.}
    \label{fig:targetlisttoonen}
\end{figure}

In TTS, higher-order dynamics are much more significant than wide hierarchical triples. Following the treatment in \cite{toonen2020}, we use the masses of the three stars and the orbital parameters to predict the possible dynamical interactions in TTS. We found that most of the systems lie in the octuple regime (Fig. \ref{fig:targetlisttoonen})
, in which the inner-binary can undergo inclination flips. Three systems are seen to lie in
the semi-secular region and are prone to stellar collisions. But the errors on
the positions are too high to be conclusive and hence more observations are needed.
\section{Conclusion}
Triple systems are now easier to probe with DEBs using high-precision
photometry and high-resolution spectroscopy. The accurate
parameters of these systems enable us to test scenarios of triple-star evolution
 and are  good checkpoints for future models of multiple star evolution. The accurate mass distributions and orbital architectures provide a platform to test and improve various theories of
multiple star formation. 
\acknowledgements{A.M. acknowledges the support provided by the Polish National Science Centre (NCN) with the grant 2021/41/N/ST9/02746. A.M., F.M., T.P., and M.K. are supported by NCN through grant no. 2017/27/B/ST9/02727.}

\bibliographystyle{ptapap}
\bibliography{moharana}

\end{document}